\documentclass{article}

\usepackage[final]{mlncp_2023}
\usepackage[utf8]{inputenc} 
\usepackage[T1]{fontenc}    
\usepackage{hyperref}       
\usepackage{url}            
\usepackage{booktabs}       
\usepackage{amsfonts}       
\usepackage{nicefrac}       
\usepackage{microtype}      
\usepackage{xcolor}         
\usepackage{graphicx}

\title{Scaling-up Memristor Monte Carlo with magnetic domain-wall physics}

\author{%
Thomas Dalgaty$^{1}$\thanks{Contact: \texttt{thomas.dalgaty@cea.fr} and \texttt{tatsuo.shibata@tdk.com}} \quad Shogo Yamada$^{2}$ \quad Anca Molnos$^1$ \quad Eiji Kawasaki$^3$ \\ \quad \textbf{Thomas Mesquida}$^1$
\quad \textbf{François Rummens}$^1$ \quad \textbf{Tatsuo Shibata}$^{2*}$ \quad \textbf{Yukihiro Urakawa}$^2$ \\ \quad \textbf{Yukio Terasaki}$^2$ \quad \textbf{Tomoyuki Sasaki}$^2$ \quad \textbf{Marc Duranton}$^3$ \\
$^1$CEA-List, Grenoble, France \quad $^2$TDK corporation, Tokyo, Japan \quad $^3$CEA-List, Paris, France
}

\begin{document}

\maketitle

\begin{abstract}
    By exploiting the intrinsic random nature of nanoscale devices, Memristor Monte Carlo (MMC) is a promising enabler of edge learning systems. However, due to multiple algorithmic and device-level limitations, existing demonstrations have been restricted to very small neural network models and datasets. We discuss these limitations, and describe how they can be overcome, by mapping the stochastic gradient Langevin dynamics (SGLD) algorithm onto the physics of magnetic domain-wall Memristors to scale-up MMC models by five orders of magnitude. We propose the push-pull pulse programming method that realises SGLD \textit{in-physics}, and use it to train a domain-wall based ResNet18 on the CIFAR-10 dataset. On this task, we observe no performance degradation relative to a floating point model down to an update precision of between 6 and 7-bits, indicating we have made a step towards a large-scale edge learning system leveraging noisy analogue devices.
    
\end{abstract}

\section{Introduction}
\label{sec:intro}

Autonomous machine learning algorithms running on distributed edge systems may open up a variety of exciting future applications and services. Before such systems may become a reality however, a diverse set of challenges must be overcome. This is not only limited to algorithmic issues, such as the scarce access to labels in a continuously evolving data stream [\citet{fahy2022scarcity}] or privacy concerns [\citet{wang2020security}], but also to hardware limitations - notably severe constraints in on-chip memory and power supply [\citet{christensen20222022}]. With regards to the latter, a potential solution may be the use of arrays of non-volatile programmable conductance elements (i.e., Memristors) to reduce the cost of repeated memory accesses when evaluating the inner-product operation in neural networks [\citet{markovic2020physics,ielmini2021brain}]. Despite a number of circuit and system level challenges [\citet{hung2021challenges}], arrays of Memristors may allow for a significant reduction in energy relative to GPUs and CPUs by performing multiplication 
in-memory, with high parallelism and in the analogue domain.
While it is hoped that weight updates derived from diverse gradient calculation or gradient approximation methods such as backpropagation [\citet{rumelhart1986learning}], equilibrium propagation [\citet{scellier2017equilibrium}] or eligibility propagation [\citet{bellec2020solution}] can be applied \textit{in-situ} directly to Memristors during the course of a learning algorithm, it in fact often proves challenging to implement these updates with a sufficiently high precision to learn a high performance model.  
The intrinsically physical and random nature of non-volatile memory, for example filamentary [\citet{ambrogio2014statistical}], phase-change [\citet{wong2010phase}] or magnetic spin-based [\citet{shibata2020linear}] technologies, impose a limit on how precisely a calculated gradient can be translated into a conductance change on a target device. 
Implementations of deterministic learning algorithms, for example stochastic gradient descent, using Memristors in this fashion typically incurs a performance degradation relative to an equivalent floating-point model [\citet{li2018efficient}].

\begin{figure}
  \centering
  \includegraphics[width=\textwidth]{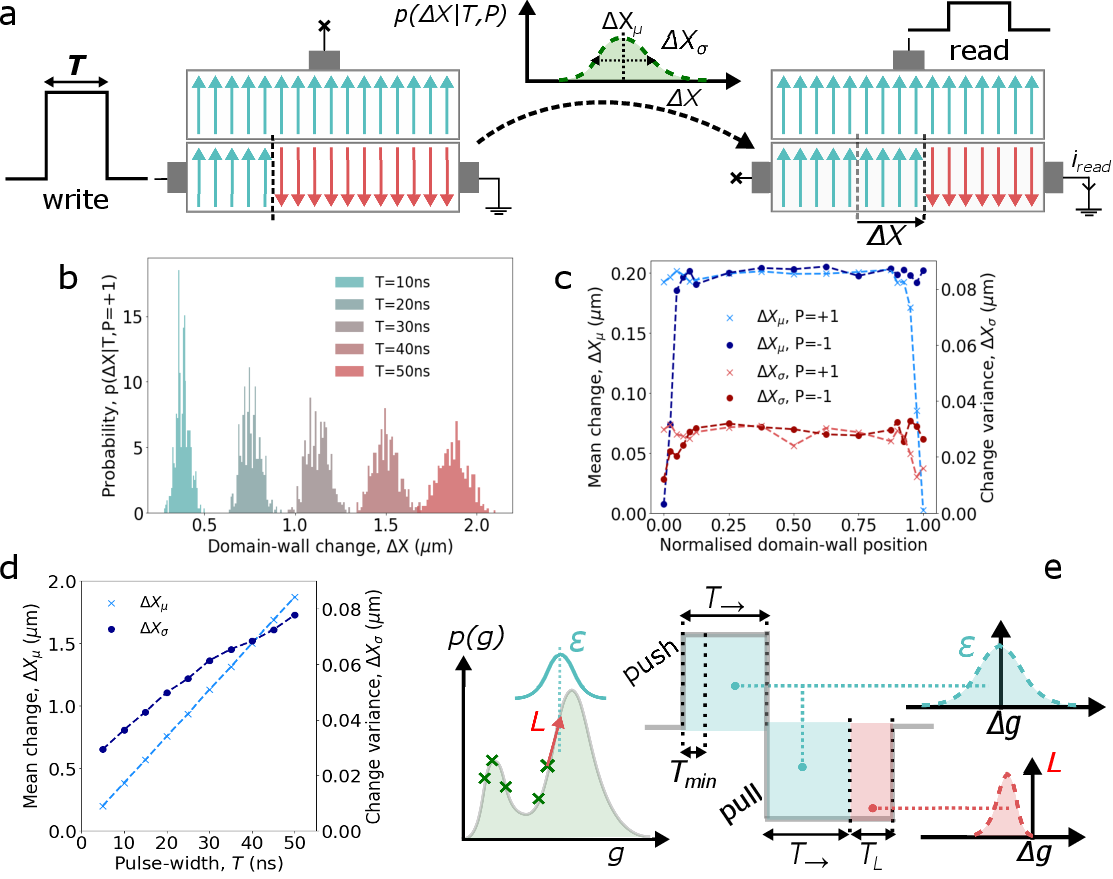}
  \caption{\textbf{SGLD with magnetic domain-walls}. (a) Domain-wall Memristor programming - domain-wall position (vertical black lines) before and after a write operation shown on the left and right. Vertical arrows (upward and downward) denote parallel and anti-parallel spin. Write and read current pulses are drawn as rectangles. (b) Simulated domain-wall position change histograms (500 trials) for varying pulse-width, $T$. (c) The independence of the mean (blue) and variance (orange) and the domain-wall position (normalised to the device length) for positive (cross) and negative (point) polarity pulses of $T=5ns$. (d) Effect of pulse-width on the mean (crosses) and variance (points) of the Memristor probability distribution. (e) (left) The principle of SGLD. Stored samples are green crosses, the Langevin gradient is a red arrow and the noise a blue Gaussian distribution. (right) A push-pull current pulse pair is shown alongside the distributions resulting from the symmetric component (blue) and the asymmetry (red).}
  \label{fig:domain_wall}
\end{figure}

An emerging alternative view however is that, by discretizing conductances into separable levels, we may be masking the rich underlying physics of Memristors. Would it not be, as is perhaps the case in our own nervous systems, preferable to find ways to compute using these phenomena instead? This idea is gradually taking hold and has lead to innovations whereby Memristor physics have been used to solve problems such as integer factorisation, Bayesian reasoning and random network generation [\citet{borders2019integer,kaiser2022hardware,dalgaty2023neuromorphic}]. They have also been directly mapped onto probabilistic machine learning algorithms such as Markov chain Monte Carlo (MCMC) sampling [\citet{dalgaty2021situ}]. In this approach, which we will refer to as \textit{Memristor Monte Carlo} (MMC), rather than viewing a device as a discrete low-precision variable, Memristors are considered to be probability distributions. Beyond the utility of allowing neural networks to better model their uncertainty [\citet{palacci2018scalable}], MCMC offers a grounded framework within which these physical probability distributions can be manipulated.
Upon programming, the conductance assumed by a Memristor is a sample from this distribution - the properties of which may be determined as a function of the programming conditions. In the case of filamentary technologies [\citet{dalgaty2021situ,dalgaty2021ex}], the mean or standard deviation of the Memristor's Gaussian distribution are defined by the programming current magnitude. 
MMC however has thus far been applied to very small neural networks and datasets - severely limiting its applicability. The reasons for this, detailed in section \ref{sec:issues}, are both algorithmic and consequences of device physics specific to filamentary Memristors. We describe in section \ref{sec:method} how, through a better choice of MCMC algorithm and Memristor technology, MMC can overcome these issues and be scaled-up massively. Specifically, we propose to leverage the physics of a three-terminal magnetic domain-wall Memristor [\cite{shibata2020linear}] (Fig.\ref{fig:domain_wall}(a)) that is currently under development, with a technique we refer to as push-pull pulse programming. In section \ref{sec:results} we demonstrate that, using this technique, domain-wall MMC may be scaled up to a dataset and model that are several orders of magnitude greater than previous demonstrations, using a standard image classification task.

\begin{figure}
  \centering
  \includegraphics[width=\textwidth]{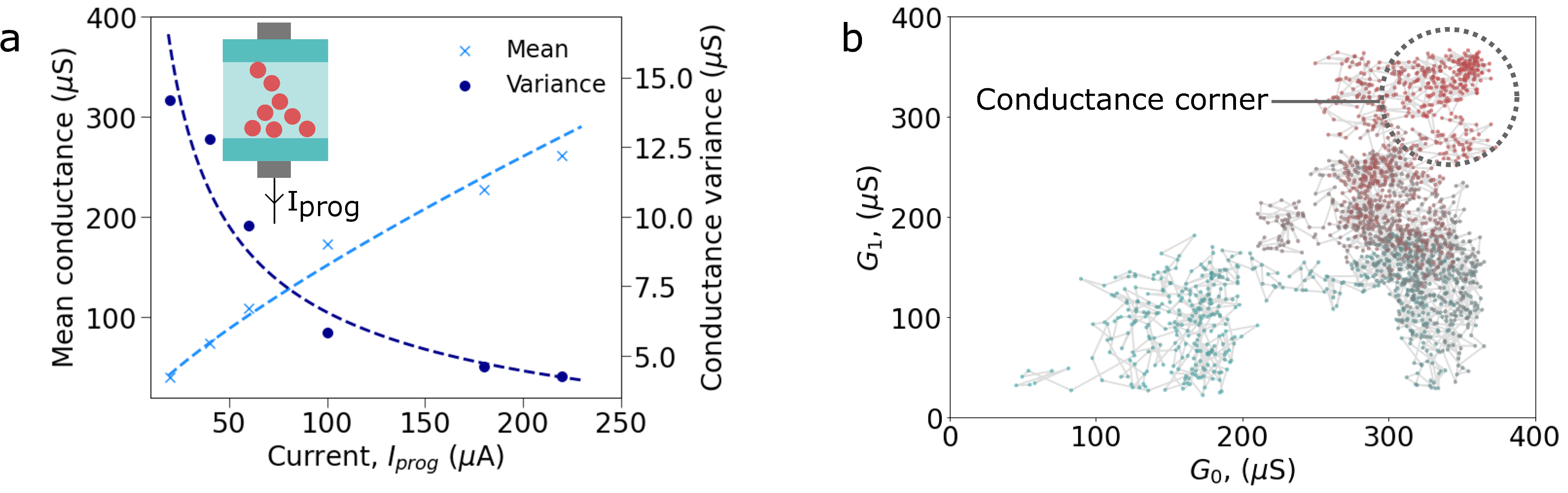}
  \caption{\textbf{Issues facing filamentary technologies for Memristor Monte Carlo}. (a) Data measured from a filamentary device [\citet{dalgaty2021situ}]. The mean (light blue) and variance (dark blue) of the Memristor probability distribution are plotted as a function of programming current. A filamentary device is shown as an inset. (b) Monte Carlo samples (points) of two conductances, $G_1$ and $G_2$, illustrate the high conductance corner issue. Point colour shows sample order (blue to red).}
  \label{fig:oxram_limits}
\end{figure}

\section{What has been limiting Memristor Monte Carlo?}
\label{sec:issues}

Despite an appealing algorithm to device pairing, previous implementations of MMC have not achieved an acceptable performance when scaled-up to large models and datasets. Principally, there are four reasons for this. The first is purely algorithmic, while the other three are related to the physical properties of filamentary Memristors:

\begin{enumerate}
\item \textbf{Poor scalability of Metropolis-Hastings} - The MCMC algorithm applied in [\citet{dalgaty2021situ}] is the Metropolis-Hastings algorithm [\citet{hastings1970monte}]. The principle of Metropolis-Hastings is to sample a proposed model based on a previous one using, typically, a Gaussian proposal distribution. The ratio of log-likelihoods and log-priors between the proposed and previous models determines a probability of accepting this proposal which is otherwise rejected.  
However, this random-walk based chain converges too slowly to high probability regions of high-dimensional posterior distributions to be practical in the context of large neural networks. In addition, the log-likelihood of each sample must be calculated with respect to the entire dataset such that, for even modest dataset sizes, the total calculation requirements over many training iterations becomes excessive. 
  
\item \textbf{Dependent mean and variance} - The general principle of filamentary Memristor technologies is that, under the application of a sufficiently large voltage pulse, a conductive filament is formed within a thin insulating layer (of some nanometers) between two electrodes. The programming current which flows during this pulse determines the structure of the filament. The measurements plotted in Fig.\ref{fig:oxram_limits}(a) show that larger currents result in thicker filaments with a higher conductance and less variability while smaller currents produce thinner, more variable, filaments of lower conductance. As a result, the current determines not only the mean of the resulting Gaussian conductance distribution, but also its variance. In practice the variance is an important free parameter in MCMC sampling algorithms and should certainly not be defined by the conductance value of the previous sample.
  
\item \textbf{The high conductance corner} - A direct result of this dependence is the high conductance corner issue. Simply, since parameters with a high conductance sample from a tighter distribution than parameters with a lower conductance, a bias exists that pushes the Markov chain to increasingly higher conductance values. In Fig.\ref{fig:oxram_limits}(b) we plot the Markov chain resulting from one thousand Monte Carlo samples of two conductances described using a filamentary device model. Indeed both parameters converge into a high conductance corner from which it is difficult to escape and, in practice, may prevent the posterior from being correctly modelled. 
  
\item \textbf{High variance to range ratio} - While MCMC methods are rooted in the generation of random samples from probability distributions, it is not the case that any distribution will do. Often for an MCMC sampler to approximate the true posterior, the variance of the proposal distribution must be tuned to lie within an optimal range of values. The range of variances of the Memristor probability distribution used in [\citet{dalgaty2021situ}] are extremely large with respect to the physical range of conductance values. At some tens of micro Siemens for example, the probability distribution contained within three standard deviations occupies over one fifth of the total conductance range.
\end{enumerate}

In order to overcome these limitations, and crucially scale-up MMC, both the MCMC method and the Memristor technology onto which it is mapped must be re-thought.

\section{Domain-wall stochastic gradient Langevin dynamics}
\label{sec:method}

Since the Metropolis-Hastings algorithm was first proposed in the 1970s, a number of modern approaches to Markov chain Monte Carlo sampling have been developed [\citet{welling2011bayesian,girolami2011riemann,chen2014stochastic,zhang2019cyclical,zhang2022low,javanbakhat2023probabilistic}] such that MCMC can now be applied to deep learning sized problems, often with state of the art results. One particularly appealing approach is the stochastic gradient Langevin dynamics (SGLD) algorithm [\citet{welling2011bayesian}]. Relative to Metropolis-Hastings, SGLD uses Langevin gradients to converge faster to high probability modes of the posterior while also being compatible with mini-batches of data. Note that, while the rejection steps of Metropolis-Hastings are not required, they may be introduced between proposals [\citet{roberts2002langevin}] in what is referred to as the Metropolis-adjusted Langevin algorithm. Given a conductance model $g$, the conductance updates calculated by SGLD may be described as

\begin{equation}
\label{eq:langevin}
{\Delta}g = \tau\nabla{_gL(g)} + \sqrt{2\tau}\epsilon ,
\end{equation}

whereby a sample from a unit noise distribution $\epsilon$ is summed with a Langevin gradient $\nabla{_gL(g)}$, which is defined as the derivative of the sum of the scaled log-likelihoods over a mini-batch of $d$ data-points and the log prior over the model: 

\begin{equation}
\label{eq:liklihood}
L(g) = \eta \sum_{i=1}^{d} - \log p(y_i|g) - \log p(g).
\end{equation}

The hyper-parameter $\eta$ acts to scale the contribution of the likelihood in order to compensate for the noise in the stochastic gradients. 
The parameter $\tau$, which may be interpreted as a learning rate, scales both the unit noise, $\epsilon$, and the Langevin gradient.
In fact a key feature of SGLD is that for $\tau < 1$, due to the square root operation, the noise term typically dominates over the Langevin gradient. This encourages the sampler to explore the posterior instead of settling into a single high likelihood mode.

Magnetic domain-wall Memristors (Fig.\ref{fig:domain_wall}(a)) are based on a physical phenomenon referred to as tunnelling magneto-resistance [\citet{julliere1975tunneling,miyazaki1995giant}]. Given a pair of magnetic materials separated by a thin insulating layer, the relative magnetic spin in the two layers influences the extent of electron tunnelling between them - therefore the conductance of the device. This effect is exploited in non-volatile memory technologies whereby the spin of a so-called \textit{free layer} may be programmed, via a spin-transfer or spin-orbit torque inducing current pulse, to be either parallel or anti-parallel with respect to the spin of the other \textit{pinned layer}.
A magnetic domain-wall is a special architecture of device wherein both parallel and anti-parallel spins are present in the free layer - on either side of a domain-wall. The location of the domain-wall, which can be moved in a linear and symmetric fashion through application of current pulses, allows the extent of tunnelling magneto-resistance between the two layers to be modified - thereby also changing the conductance. However, the precision of these conductance updates will ultimately be limited by the inherent randomness associated to domain-wall motion due to effects like surface roughness and thermal fluctuations [\citet{martinez2011stochastic,liu2021domain}]. 
We propose that this inherent probabilistic behaviour can be leveraged through a Memristor Monte Carlo approach. Just as the programming current magnitude determines the mean and variance of filamentary Memristor distributions (Fig.\ref{fig:oxram_limits}(a)), the pulse-width of the current applied to the free layer in a domain-wall device will determine the shape of its own conductance change probability distribution. Numerical simulations of domain-wall motion
give examples in Fig.\ref{fig:domain_wall}(b) of the domain-wall position change probability distribution due to positive polarity pulses ($P=+1$) of different pulse-widths, $T$. We observe that this distribution can be well described using a Gaussian - $p({\Delta}X|T,P) = \mathcal{N}({\Delta}X_{\mu},{\Delta}X_{\sigma})$ - with a mean, ${\Delta}X_{\mu}$, and variance, ${\Delta}X_{\sigma}$, that both increase as a function of pulse-width (Fig.\ref{fig:domain_wall}(d)).
Crucially however, with respect to filamentary technologies, the mean and variance of these distributions are independent of the position of the domain-wall. Fig.\ref{fig:domain_wall}(c) plots the mean and variance of the position change from different initial positions along the length of the device for a fixed current pulse-width ($T=5ns$). We observe that regardless of the domain-wall position, and therefore Memristor conductance, the mean and variance of the resulting distribution remain constant for positive and negative polarity pulses.
A saturation effect is observed only when the domain-wall reaches the extremities of the free layer and cannot physically be moved further. More detail on our domain-wall device model and these simulations can be found in the supplementary material.

As discussed, due to tunnelling magneto-resistance, these position changes translate directly into a conductance change distribution $p({\Delta}g|T,P)=\mathcal{N}({\Delta}g_{\mu},{\Delta}g_{\sigma})$, which can be leveraged within the context of analogue machine learning. We propose to exploit this distribution by introducing a novel \textit{push-pull} pulse programming method which, using the physics of domain-wall motion, realises the conductance update required by equation \ref{eq:langevin}. By applying an asymmetrical pair of pulses with opposing polarities, labelled \textit{push} and \textit{pull} in Fig.\ref{fig:domain_wall}(e), the domain-wall position can be moved forwards with a positive polarity pulse before being pulled backwards again with a negative polarity pulse. The resulting conductance change will be sampled from the distribution

\begin{equation}
p({\Delta}g) = \frac{1}{\sqrt{\Delta{g_{\sigma,\rightarrow}}^2 + \Delta{g_{\sigma,\leftarrow}^2}}\sqrt{2\pi}}
exp\left[{\frac{-(\Delta{g_{\mu,\rightarrow}}-\Delta{g_{\mu,\leftarrow}})^2}{2\sqrt{\Delta{g_{\sigma,\rightarrow}}^2 + \Delta{g_{\sigma,\leftarrow}}^2}}}\right],
\end{equation}

which is a result of summing the Gaussian distributions that result from the push and pull pulses, $p({\Delta}g|T_\rightarrow,P=+1)$ and $p({\Delta}g|T_\leftarrow,P=-1)$.
With respect to the update of equation \ref{eq:langevin}, the symmetrical push-pull component (shaded in blue in Fig.\ref{fig:domain_wall}(e)) serves to generate a sample from a zero-centred conductance change distribution with a variance $\sqrt{2\tau}\epsilon$. The asymmetry introduced by lengthening one of the pulses offsets this distribution by the Langevin gradient $\tau\nabla{_gL(g)}$. 
In the case of a negative Langevin gradient where the pull is lengthened with respect to the push (as in Fig.\ref{fig:domain_wall}(e)), $T_{\leftarrow}=T_{\rightarrow}+T_L$, where $T_L$ is the pulse-width that offsets $p({\Delta}g)$ by
$\tau\nabla{_gL(g)}$.
Crucially, it should be noted that because the total variance of $p({\Delta}g)$ is $\sqrt{2\Delta{g_{\sigma,\rightarrow}}^2+\Delta{g_{\sigma,L}}^2}$, where $\Delta{g_{\sigma,L}}$ is the variance incurred by $T_L$, the noise related to offsetting the domain-wall by $\tau\nabla{_gL(g)}$ may effectively be absorbed into $\sqrt{2\tau}\epsilon$ by setting $T_\rightarrow$ appropriately. In spite of Memristor variability therefore, the Langevin gradient is programmed into the Memristor precisely. Please refer to the supplementary material for a further explanation of this surprising result.
\section{Results}
\label{sec:results}

In order to evaluate 
whether our proposed domain-wall based SGLD push-pull update will indeed allow Memristor Monte Carlo to scale-up, we train a ResNet18 [\citet{he2016deep}] model on the CIFAR-10 benchmark. 
Relative to previous implementations of MMC using filamentary Memristors, this represents a three order of magnitude increase in dataset size and a five order of magnitude increase in model size [\citet{dalgaty2021situ}]. 
We assume that each parameter is defined by the subtraction of two domain-wall conductances and, by application of a scaling factor, the effective synaptic weights are mapped onto a range of -1 to +1. We place a uniform prior over the weights reflecting this - in equation \ref{eq:liklihood}, $p(g) = U(-1,+1$) and, based on the data from Fig.\ref{fig:domain_wall}, we model the conductance change distribution for $p({\Delta}g|T,P)$ as a Gaussian random variable. To account for eventual programming limitations, we introduce $T_{min}$. This is the shortest possible pulse and results accordingly in a Gaussian distribution which, at three standard deviations, is equal to a given minimum precision. For example, at 3-bits of precision, we assume that this minimum Gaussian has an effective width spanning a range of $0.25$ - i.e., $\nicefrac{1}{8}^{th}$ of the parameter range. 
More details of the training method incorporating the proposed push-pull domain-wall update can be found in the supplementary material.

In order to understand how device precision impacts domain-wall SGLD we report the effect of the number of bits of update precision (limited by $T_{min}$) as well as the number of posterior
samples in Table \ref{table:cifar10}. 
Above 6-bits we observe that, despite an increasingly variable device, an accuracy equal to that obtained by an equivalent floating-point precision model is achieved. It is in fact surprising that the approach works so well at 7-bits since the noise contributed by the symmetric component of the push-pull is in excess of the ideal value of $\sqrt{2\tau}\epsilon$.
The first drop off in performance is observed at an update precision of 6-bits. For sixty-four posterior samples, there is a loss of 4.9\% relative to the best accuracy (achieved at 9-bits). Below 6-bits, the parasitic noise due to ever increasing values of $T_{min}$ begins to overwhelm the Langevin gradient signal and, as a result, the high probability modes of the posterior are not effectively sampled from - degrading test accuracy. 
Furthermore, as the update precision reduces, we note that task performance degradation can be mitigated by storing more posterior samples. While we attempted to incorporate the filamentary MMC approach of [\citet{dalgaty2021situ}] into our comparison, we did not succeed in training a model that surpassed random guessing. This highlights the advance made by our approach relative to the current state of the art. 
Finally, we emphasise that, in contrast to deterministic Memristor-based machine learning whereby device variability limits how precisely a gradient update can be realised (see supplementary material for a comparison with a deterministic method), our method permits the Langevin gradient update to be realised precisely. This is not only because the variability incurred by the asymmetry of the push-pull is absorbed into the symmetrical component but also the fact that the gradient value is coded by a potentially fine difference between push and pull pulse-widths. Rather, it is the lower-bound on the variance of the noise distribution, dictated by $T_{min}$, that limits performance. It is therefore important to understand how this quantity can be minimised in domain-wall Memristors under development.

\begin{table}
  \caption{Accuracy, as a percentage, of domain-wall MMC on the CIFAR-10 test set over a range of posterior samples (columns). The left-most column denotes the update precision.}
  \label{table:cifar10}
  \centering
  \begin{tabular}{lllllll}
    \toprule
    Precision     & 64 & 32 & 16 & 8 & 4 & 2 \\
    \midrule
    32-bit FP       & 86.3 & 86.1 & 85.6 & 84.9 & 84.5 & 83.9 \\
    10-bit          & 86.2 & 85.8 & 85.4 & 84.9 & 84.3 & 83.7 \\
    9-bit           & 86.5 & 86.3 & 85.7 & 85.1 & 84.5 & 84.3 \\
    8-bit           & 86.0 & 85.6 & 85.4 & 85.0 & 84.1 & 84.0 \\
    7-bit           & 85.7 & 85.6 & 85.1 & 85.1 & 83.4 & 83.1 \\
    6-bit           & 81.6 & 81.2 & 80.1 & 80.0 & 78.4 & 76.3 \\
    5-bit           & 70.1 & 70.0 & 69.9 & 69.0 & 67.1 & 65.4 \\
    4-bit           & 51.6 & 51.2 & 50.1 & 48.2 & 47.7 & 42.3 \\
    \bottomrule
  \end{tabular}
\end{table}

\section{Conclusion}
\label{sec:conclusion}

We have proposed the \textit{push-pull} update which maps the stochastic gradient Langevin dynamics algorithm onto the physics of magnetic domain-wall Memristors. We applied this update to train a ResNet18 model, finding that, even with imprecise and noisy conductance updates, our results were comparable to those of a floating-point precision model. Crucially, our method promises that Memristor Monte Carlo can be scaled-up - here we used a model and a dataset five and three orders of magnitude larger respectively than the current state of the art. Our next steps will be to work towards a physical system using fabricated domain-wall Memristors that execute MMC to achieve on-chip adaptation and enable edge learning. Insights as to what this future system might resemble may be gained by looking towards intriguing ideas in neuroscience which discuss how human brains,which are also subject to intrinsic noise, may too compute using Monte Carlo methods [\citet{aitchison2016hamiltonian,echeveste2020cortical}].

\section{Acknowledgements}
\label{sec:ack}
We would like to thank our colleagues Kazuki Nakada, Toru Oikawa and Toshiki Gushi (TDK) as well as Djohan Bonnet, Renaud Mourot, Frédéric Surleau, Philippe Doré and Cédric Auliac (CEA) for their support in preparing this work. We would also like to thank the TDK corporation and Horizon Europe (dAIedge, Grant Agreement Number: 101120726) for financing this work.

\bibliography{mmc_main}

\newpage

\section{Supplementary material}
\label{sec:appendix}

\subsection*{Detail on the learning approach}

In our implementation of Memristor Monte Carlo we took inspiration principally from the paper which originally proposed stochastic gradient Langevin dynamics [\citet{welling2011bayesian}]. However, instead of decaying $\tau$ from an initial value, we keep it fixed for the entire learning procedure. This is something we observe is often done in practice in the literature. The optimal value for $\tau$ across the studied range of update precision was found to be ${2\times10^{-5}}$ for a batch-size of size 48 and where $\eta=1\times10^5$. 

We assume that all convolutional and fully-connected layer parameters are realised by two domain-wall Memristors. The effective parameter is the subtraction of the conductances of these two devices. When the parameter is positive we assume that only the positive device has its domain-wall position updated and that the negative device is programmed to its lowest conductance value. Conversely, when updating negative values we suppose only the negative device is updated and the positive device is programmed to its minimum value.

We propose that the affine transform parameters of the batch normalisation operation are performed in the digital domain, using the weighted sums calculated by the Memristor arrays. They are therefore optimised at a precision of 32-bit floating-point with a learning rate of $2\times10^2$. We believe this is a practical choice, since batch normalisation is used to scale the neuron activations after evaluation of the inner-product (calculated by the Memristor array) and it is not clear what the advantage of doing this with Memristors would in fact be. In any case, the number of batch normalisation parameters are relatively few compared to the large number of convolutional and fully-connected parameters in ResNet18.

While typical digital MCMC approaches may collect a large number of samples which are thereafter thinned, this is not feasible in a Memristor based system which may only have a capacity to store $N$ samples from the posterior distribution of the model. In an eventual system based on Memristors, such a post-training thinning may need to be replaced by an \textit{online thinning} procedure. This is what was implemented in our experiments. In [\citet{dalgaty2021situ}] (where no thinning procedure is considered) after a sample was accepted for a model sample at index $n$, the algorithm proceeds immediately onto propose a model sample at index $n+1$. A sample index may, for example, refer to a collection of Memristor arrays that store one full version of the model. Each such collection is referred to as an index. However, for a large model like a ResNet18,
storing successive and highly correlated samples in this fashion risks to provide a poor description of the posterior - in particular if we are constrained in how many Memristors are at our disposal. In our simulations we therefore incorporate an online thinning strategy whereby, after accepting a model sample at index $n$, the algorithm will proceed to generate new models upon each successive mini-batch of data at the same sample index $n$. Unlike [\citet{dalgaty2021situ}], we do not pass directly onto the next sample index, $n+1$. We only pass to index $n+1$ - leaving model $n$ encoded by a collection of Memristors to be used later in inference - after a \textit{thinning downcounter} has been decremented from an initial value down to 0. The thinning downcounter is set to an initial value $C_{th}$ and then decremented by one after generating a new model
upon each mini-batch. 
When the downcounter is equal to zero the most recently proposed model is stored definitively at index $n$, and the algorithm proceeds to generate new models at sample index $n+1$, and so on. In our experiments, $C_{th}=521$. Furthermore, in order to store samples from diverse regions of the posterior, we add a further thinning mechanism whereby this downcounter is only decremented during a window at the end of a cycle of $S$ mini-batches. 
Although we do not use a cosine schedule, this was partially inspired by [\citet{zhang2019cyclical}] where samples are collected at the end of a cosine schedule after having, ideally, converged to different high probability mode from the previous cycle.
We use a cycle length of $S=20,840$ and only decrement the downcounter after sample $18,756$ - thereby recording four samples per cycle. Samples were collected over $500$ epochs. The test accuracy reported in the main text results from an inference with the most recently recorded samples. The inference itself computes the average softmax distribution of all stored samples and assigns a classifcation based on the peak probability. Both the domain-wall Memristor based and the floating-point models were trained with the same set of hyperparameters.

PyTorch was used to implement the algorithm and the Langevin gradients were calculated for each parameter using the autograd module - no regularisation methods or momentum were used. The ResNet implementation was taken from the Torchvision package and the initial parameters were randomly initialised within the range -1 to +1. The CIFAR-10 database was also taken from Torchvision and each image was normalised before being input into the first convolutional layer with the recommended mean and standard deviation for the CIFAR-10 dataset.

\subsection*{Detail on the magnetic domain-wall simulations}

The one-dimensional Landau-Lifshitz-Gilbert equation was used to calculate the current-driven domain-wall motion [\citet{emori2013current}]. It can be represented by two coordinates, the domain-wall position $X$ and the domain-wall phase $\phi$ where

\begin{equation}
(1+\alpha^2)\frac{dX}{dt} = (1+\alpha\beta)\frac{\mu_BP}{eM_s}j+\frac{\gamma_0{\Delta}H_K}{2}sin(2\phi)+\alpha\gamma_0\Delta(H_p+H_{th}) ,
\end{equation}

and 

\begin{equation}
(1+\alpha^2)\frac{d\phi}{dt} = (\beta-\alpha)\frac{\mu_BP}{eM_s\Delta}j-\alpha\frac{\gamma_0H_K}{2}sin(2\phi)+\gamma_0(H_p+H_{th}).
\end{equation}

Here $\alpha$ is the Gilbert damping parameter, $\beta$ is the parameter defining strength of non-adiabaticity, $\mu_B$ is Bohr magnetron, $P$ is the spin polarization, $e$ is the electron charge, $M_s$ is the saturation magnetization, $j$ is the current density, $\gamma_0$ is the gyro magnetic ratio, $\Delta$ is the domain-wall width and $H_K$ is the anisotropy field.

A pinning effect $H_p$, which is introduced as a simple periodic function assuming irregularities in the DW layer, and a thermal fluctuation $H_{th}$ are introduced as effective magnetic fields:

\begin{equation}
H_p(X)=-\frac{1}{2\mu_0M_sL_yL_z}\frac{V_0\pi}{p}sin(\frac{2{\pi}X}{p}) ,
\end{equation}

and 

\begin{equation}
H_{th}(t) = \eta(t)\sqrt{\frac{2{\alpha}K_BT}{\gamma_0\mu_0M_sL_yL_z{\Delta}dt}}.
\end{equation}

Here, $V_0$ is the energy barrier between adjacent minima of the pinning profile, $p$ is the spatial periodicity, $K_B$ is the Boltzmann constant, $\mu_0$ is the vacuum permeability and $\eta$ is a function that generates unit Gaussian distributed random numbers.

In our simulations we have used $\alpha=0.07$ and $\beta=0.06$ for the domain-wall layer by assuming a [Co/Pd]-based multilayer system with perpendicular magnetic anisotropy (PMA) [\citet{sajitha2010magnetization}]. $P$, $M_s$ and $\Delta$ were assumed to be $0.55$, $8\times10^5$A/m and $5$nm, respectively. Since the values of $H_K$, $V_0$, $p$ and $\gamma_0$ are not yet known for this system, values from a previous study of one-dimensional simulation in a PMA film were used as in [\citet{martinez2011stochastic,emori2013current}]. The dimensions of the domain-wall layer are $L_y=60$nm in width and $L_z=7.5$nm in thickness. The temperature $T$ and the temporal time step $dt$ were set to 300K and 1ps respectively.

In the simulations which produced the results for Fig.\ref{fig:domain_wall}(b) and (d), the pulse current density was fixed at $|j|=1\times10^{12}A/m^2$, and only the pulse width was swept in 5ns steps from 5ns to 50ns. This was performed for both pulse polarities - positive and negative. For each pulse condition, the domain-wall position before and after the application of the pulse was recorded as it progressed along the length of the device. This was repeated five-hundred times, allowing us to characterise the resulting distributions.

For the result of Fig.\ref{fig:domain_wall}(c), we performed a similar simulation where the pulse current density was once again set to $|j|=1\times10^{12}A/m^2$ but the pulse-width was fixed at 5ns. A boundary condition was introduced at the longitudinal end of the device whereby the movable region of the domain-wall was limited, 0 < $X$ < $L_{DW}$. To generate each point on Fig.\ref{fig:domain_wall}(c), one hundred positive and negative pulses were applied over a range of initial domain-wall positions $X_0$, between 0 and $L_{DW}$. The results are normalised for $L_{DW}=4{\mu}m$.

\subsection*{Comparison with SGD using push-pull updates}

While we do not advocate that our proposed push-pull updates are a suitable method to realise deterministic conductance updates to Memristors, we still find the comparison indicative of the the ability of domain-wall SGLD to maintain a high test accuracy as the device precision reduces. In these experiments we set the symmetric component of the push-pull equal to the minimum pulse-width $T_{min}$. This is because the noise that it generates is no longer viewed to be integral part of the update but in fact as a parasitic contribution. The performance of this method was found to be optimal when using a learning rate equal to $4\times10^{-2}$. The rest of the hyper-parameters are consistent with those described for the standard SGLD updates. The test accuracy over a range of update precisions are reported in Table \ref{table:sgd} and compared with the accuracy of our propsed domain-wall SGLD method.

\begin{table}
  \caption{Accuracy of a push-pull pulse pair based stochastic gradient descent update compared to the stochastic gradient Langevin dynamics result for sixty-four stored samples.}
  \label{table:sgd}
  \centering
  \begin{tabular}{lll}
    \toprule
    Precision   &  DW-SGD & DW-MMC  \\
    \midrule
    10-bit    & 85.1 & 86.3 \\
    9-bit     & 82.1 & 86.6 \\
    8-bit     & 75.0 & 86.0 \\
    7-bit     & 61.1 & 85.7 \\
    6-bit     & 47.8 & 81.6 \\
    5-bit     & 26.5 & 70.1 \\
    4-bit     & 10.1 & 51.6 \\
    \bottomrule
  \end{tabular}
\end{table}

The results show an increasingly poor test accuracy as the update precision reduces - before a collapse below 6-bits of precision. The accuracy obtained by domain-wall SGLD (corresponding to the values reported in the main paper for 64 posterior samples) is shown alongside. While at 10-bits the SGD actually obtains an impressive performance, very close to that of MMC, the test accuracy quickly diminishes as the precision is reduced further. In particular at a precision of 7-bits and below there is a vast difference between SGD and our proposed method.

\subsection*{More detail on the noise-free Langevin gradient update}

\begin{figure}
  \centering
  \includegraphics[width=\textwidth]{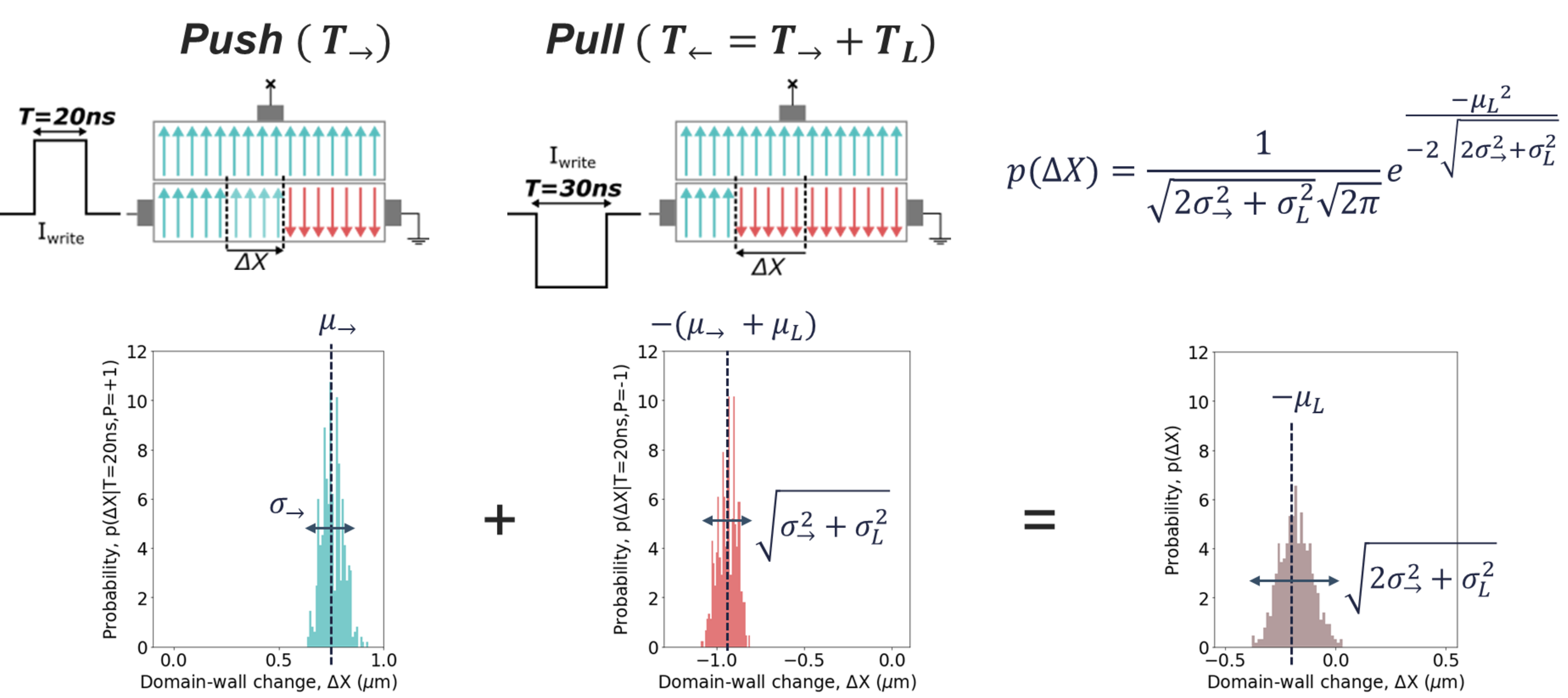}
  \caption{An example of the distributions and their sum resulting from the push and pull programming operations.}
  \label{suppfig:noiseless}
\end{figure}

If conductance variability is an intrinsic and unavoidable fact when working with Memristors, then how can it be that we achieve a noise-free Langevin gradient domain-wall conductance update ? In reality there is no free-lunch - the pulse-width $T_L$ does in fact incur a parasitic noise $\Delta{g_{sigma,L}}$ that increasses with pulse duration as observed in Fig.\ref{fig:domain_wall}(d). However, the SGLD update we aim to implement in our Memristor requires a second element - a zero-centred Gaussian sample. The variance of this sample is determined by (for the case of a negative gradient as in Fig.\ref{suppfig:noiseless}) the push pulse, $T_\rightarrow$. The longer this pulse-width, the higher the variance of the zero-centred Gaussian component. Since, as often happens to be the case in nature, our conductance change distributions are Gaussian we know that, according to the Gaussian summing law, the total variance incurred in the push and the pull is $\sqrt{2\sigma_\rightarrow^2+\sigma_L^2}$. Therefore, in order to obtain the desired variance $\sqrt{2\tau}\epsilon$, we will need to define $T_\rightarrow$ such that, when added to the variance incurred by $T_L$, they sum to the this value. This may be also understood as \textit{shortening} $T_\rightarrow$ by a time defined by the variance incurred by the $T_L$. By determining the pulse-width $T_\rightarrow$ to achieve a desired total variance (when summed with the variance incurred by $T_L$), we can imagine that we are effectively absorbing the noise incurred in the application of the Langevin gradient into the noise we in any case require to draw a sample with a variance of $\sqrt{2\tau}\epsilon$. 

However it is important to note that this in only true between a lower and upper limit. The lower limit, as discussed in the main body of the paper, is determined by a minimum pulse-width $T_{min}$ which gives rise to the narrowest possible Gaussian distribution. There also however exists an upper limit. This occurs when the desired Langevin gradient is sufficiently large that the variance incurred due to a pulse duration of $T_L$ exceeds the desired variance value of $\sqrt{2\tau}\epsilon$. In practice, we have not found exceeding the upper limit to be problematic on the CIFAR-10 task. Intuitively, if the gradient is very large, it suffices to take a large step in the correct direction instead of realising this large gradient very precisely in the conductance change of the Memristor.

\end{document}